\documentclass{lmcs}

\usepackage{hyperref}
\usepackage{graphicx}
\usepackage{amsmath, amssymb}
\usepackage{array}
\usepackage{braket}
\usepackage[section]{placeins}
\usepackage{orcidlink}

\begin{document}

\title[BB84-Simulated Hybrid Encryption]{\texorpdfstring{Quantum-Classical Hybrid Encryption Framework Based on Simulated BB84 and AES-256: Design and Experimental Evaluation}{Quantum-Classical Hybrid Encryption Framework}}

\author[H.~Mozo]{Hector~E.~Mozo\,\orcidlink{0009-0008-2426-623X}}
\address{Department of Computer Science, University of the People, California, USA}
\email{HectorMozo@my.uopeople.edu, hectormozo308@gmail.com}

\begin{abstract}
This paper presents the design, implementation, and evaluation of a hybrid encryption framework that combines quantum key distribution, specifically a simulated BB84 protocol, with AES-256 encryption. The system enables secure file encryption by leveraging quantum principles for key generation and classical cryptography for data protection. It introduces integrity validation mechanisms, including HMAC verification and optional post-quantum digital signatures, ensuring robustness even in the presence of quantum-capable adversaries. The entire architecture is implemented in Python, with modular components simulating quantum key exchange, encryption, and secure packaging. Experimental results include visual testing of various attack scenarios, such as key tampering, HMAC failure, and file corruption, demonstrating the effectiveness and resilience of the approach. The proposed solution serves as a practical foundation for quantum-aware cybersecurity systems.
\end{abstract}

\keywords{Quantum Cryptography; Quantum Key Distribution (QKD); Simulated Quantum Protocols; BB84 Protocol; Hybrid Encryption; AES-256; Post-Quantum Security; HMAC Validation; Modular Cryptographic Architectures; Cryptographic Metrics; Scientific Visualization; Educational Cryptography Tools}

\maketitle

 \section{Introduction}
As quantum computing evolves from theoretical promise to technological reality, the urgency for quantum-resistant security solutions grows accordingly~\cite{mosca2018cybersecurity}. Classical encryption schemes such as AES-256, though currently secure, are vulnerable to future quantum adversaries equipped with algorithms like Grover’s and Shor’s~\cite{grover1996fast,shor1994algorithms}. In this context, hybrid cryptographic systems that combine classical robustness with quantum principles offer a promising path forward~\cite{hughes1997quantum}.

One of the most prominent quantum key distribution (QKD) protocols is BB84, which exploits quantum states of photons to enable secure key exchange~\cite{bennett1984quantum}. Although real-world quantum channels are limited by cost and infrastructure, BB84 can be simulated with high fidelity using quantum computing frameworks like IBM Qiskit~\cite{abraham2019qiskit}. When paired with robust classical cryptography, such as AES-256, a hybrid system can achieve both forward secrecy and post-quantum resilience~\cite{ekert1991quantum}.

This paper introduces a Python-based encryption application that implements a hybrid cryptographic framework using a simulated BB84 protocol for quantum key generation and AES-256 for data encryption. To ensure message integrity and protect against tampering, the system includes a hash-based message authentication code (HMAC)~\cite{bellare1997hmac} and optional post-quantum digital signature support using CRYSTALS-Dilithium~\cite{schwabe2018crystals}. The architecture is modular, fully implemented in Python, and includes a graphical interface that visualizes the quantum key distribution process.

The proposed system is evaluated through extensive visual test cases, including successful and adversarial scenarios, such as key mismatches, HMAC failures, and encrypted file corruption. The results demonstrate the viability of using quantum-inspired mechanisms within practical classical infrastructures, especially for educational, research, and security-focused applications.

\section{Related Work}
The convergence of quantum key distribution (QKD) and classical cryptography has received increasing attention in recent years as a response to the emerging threats posed by quantum computing. One study proposed a hybrid encryption system leveraging QKD and symmetric encryption to secure medical data in telemedicine networks, highlighting the potential of quantum-classical integration for real-world applications~\cite{shinde2024hybrid}. Such systems enhance confidentiality by using quantum-generated keys for conventional encryption algorithms like AES.

In the context of practical experimentation, several research initiatives have focused on simulating the BB84 protocol using quantum computing frameworks. Open-source platforms such as IBM Qiskit have enabled the emulation of photon polarization, basis selection, and measurement outcomes, thereby making QKD accessible in software environments~\cite{kairost_bb84_sim}. These simulations serve as powerful educational tools and testbeds for conceptual validation when actual quantum hardware is unavailable. Moreover, the theoretical security of BB84 has been formally established through rigorous proofs in the literature~\cite{shor2000proof}.

Post-quantum cryptography (PQC) is also being explored as a classical alternative resistant to quantum attacks. Algorithms based on lattice problems, such as CRYSTALS-Dilithium, are leading candidates for future digital signature standards. Studies have proposed incorporating these into secure data transmission and storage systems, particularly in cloud and blockchain-based infrastructures~\cite{liu2022postquantum}.

Compared to existing approaches, the system proposed in this paper uniquely integrates a simulated BB84-based key exchange protocol with AES-256 file encryption, enhanced by HMAC-based integrity checking and optional post-quantum digital signatures. The entire implementation is modular and includes a graphical interface for visualization and real-time interaction. Additionally, the framework has been validated using visual testing across multiple adversarial scenarios, offering a robust educational and experimental platform.

\begin{table}[ht]
\centering

\caption{Comparative Analysis of Related Quantum-Classical Security Approaches}
\label{tab:related_qc_approaches}
\footnotesize
\begin{tabular}{|>{\raggedright\arraybackslash}p{2.5cm}|>{\raggedright\arraybackslash}p{2.5cm}|>{\raggedright\arraybackslash}p{2.5cm}|>{\raggedright\arraybackslash}p{2.5cm}|>{\raggedright\arraybackslash}p{2.5cm}|}
\hline
\textbf{Study / System} & \textbf{Cryptographic Strategy} & \textbf{Execution Environment} & \textbf{Functional Scope} & \textbf{Identified Limitations} \\ \hline
Hybrid QKD for Medical IoT~\cite{shinde2024hybrid} & QKD + Symmetric Encryption & Theoretical quantum channel & Securing medical imaging in telemedicine & Depends on unavailable quantum infrastructure \\ \hline
BB84 Simulation (Qiskit)~\cite{kairost_bb84_sim} & Software-based BB84 QKD simulation & IBM Qiskit (classical hardware) & Educational and conceptual demonstration & Lacks integration with operational encryption workflows \\ \hline
PQC in Cloud \& Blockchain~\cite{liu2022postquantum} &  Lattice-based Post-Quantum Signatures (e.g., Dilithium) & Classical cloud infrastructure & Secure storage and decentralized communication & No real-time quantum interaction or visualization \\ \hline
BB84 Security Proof~\cite{shor2000proof} & Theoretical Security Model (Quantum Information Theory) & Mathematical framework only & Formal security justification of BB84 & Not implemented or evaluated under practical constraints \\ \hline
Proposed Framework (This Study) & Simulated BB84 + AES-256 + HMAC + Dilithium Signature & Modular Python (software-only) & Full hybrid encryption, integrity, and GUI & Requires manual validation of key B and secure user workflow \\ \hline
\end{tabular}

\vspace{0.3cm}
\emph{A structured comparison of related systems is presented in Table~\ref{tab:related_qc_approaches}, highlighting the technical differentiators of the proposed framework relative to prior QKD, simulation, and post-quantum approaches.}
\end{table}

\section{Methodology}
The encryption workflow is structured as a layered system of cryptographic and quantum simulation algorithms, each mapped to an independent module. These components collectively implement a hybrid cryptographic pipeline that enables secure file encryption based on simulated quantum key distribution (BB84), symmetric AES-256 encryption, HMAC-based key integrity verification, and optional post-quantum digital signatures. This section details the internal mechanisms and interactions among these algorithmic components.

\begin{figure}[ht]
\caption{Illustrates linear activation and real-time monitoring of each module, including decryption coordination.}
\label{fig:architecture}
\centering
\includegraphics[width=\linewidth]{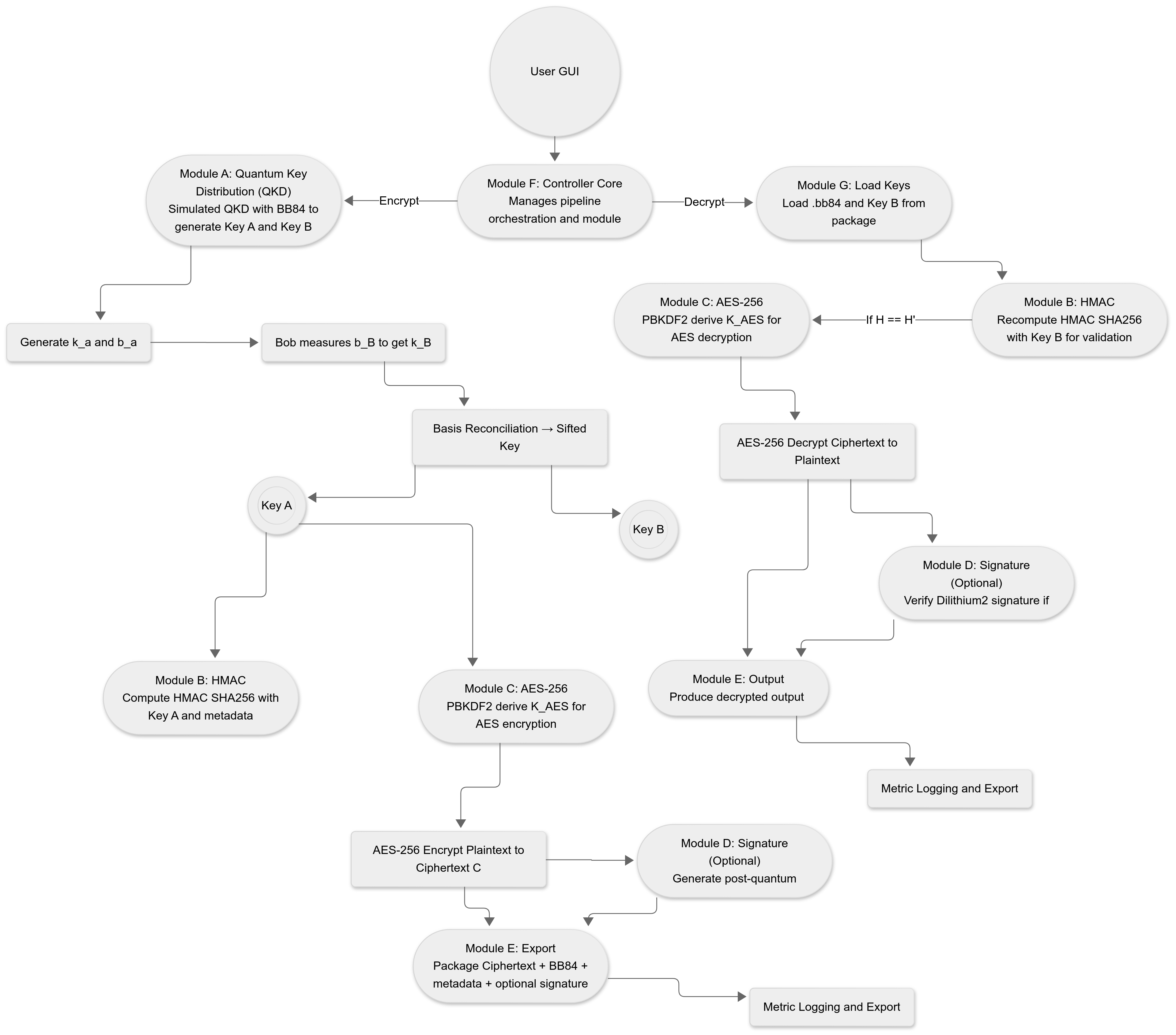}

\end{figure}

\FloatBarrier

\subsection{Quantum Key Distribution via BB84 Simulation ---Figure~\ref{fig:architecture}} (Module A)

Key generation begins by simulating the BB84 quantum protocol using Qiskit's Aer simulator. Each bit of the plaintext key is encoded into a quantum state using one of two bases selected at random:

\begin{itemize}
    \item \textbf{Z-basis}: $\left\{ \vert 0 \rangle, \vert 1 \rangle \right\}$
    \item \textbf{X-basis}: $\left\{ \vert + \rangle = \frac{1}{\sqrt{2}} \left( \vert 0 \rangle + \vert 1 \rangle \right),\ \vert - \rangle = \frac{1}{\sqrt{2}} \left( \vert 0 \rangle - \vert 1 \rangle \right) \right\}$
\end{itemize}

The simulation proceeds as follows:

\begin{itemize}
    \item Alice generates a random bitstring $k_A \in \{0,1\}^n$ and a basis string $b_A \in \{Z, X\}^n$.
    \item Bob selects his own basis string $b_B \in \{Z, X\}^n$.
    \item Bob measures the received qubits:
    
    \[
    k_B = \text{measure}(k_A, b_A, b_B) \tag{1}
    \]
    
    \item After basis reconciliation, they derive the sifted key:
    
    \[
    s = \left\{ k_i\ \vert\ b_{A_i} = b_{B_i} \right\} \tag{2}
    \]
    
    \item From this, Key A and Key B are obtained:
    
    \[
    K_A = s \tag{3}
    \]
    
    \[
    K_B = \text{subset}(K_A) \tag{4}
    \]
\end{itemize}

This process enforces randomness, non-cloning, and interception resistance—key principles in QKD.

\subsection{Symmetric File Encryption with AES-256 --- Figure~\ref{fig:architecture}} (Module C)

Once the sifted key is established, the system derives a secure symmetric key using PBKDF2:

\[
K_{\text{AES}} = \text{PBKDF2}(K_A, S, c) \tag{5}
\]

Where:

\begin{itemize}
    \item $S$ is a 128-bit random salt.
    \item $c$ is the iteration count (typically $10^5$).
\end{itemize}

The plaintext file $M$ is padded using PKCS\#7 and encrypted via AES in CBC mode:

\[
C = \text{AES}_{\text{CBC}}(K_{\text{AES}}, IV, M) \tag{6}
\]

Where $IV$ is a 128-bit random initialization vector.

\subsection{Integrity Verification via HMAC --- Figure~\ref{fig:architecture}} (Module B)

To validate $K_B$ without revealing $K_A$, a secure HMAC is used.

During encryption:

\[
H = \text{HMAC}_{\text{SHA256}}(K_A, \text{metadata}) \tag{7}
\]

During decryption:

\[
K'_A = \text{PBKDF2}(K_B, S, c) \tag{8}
\]

\[
H' = \text{HMAC}_{\text{SHA256}}(K'_A, \text{metadata}) \tag{9}
\]

If and only if:

\[
H = H' \tag{10}
\]

then decryption proceeds.

\subsection{Post-Quantum Digital Signature (Optional) --- Figure~\ref{fig:architecture}} (Module D)

For long-term authenticity, the system optionally signs the ciphertext with CRYSTALS-Dilithium Level 2:

\[
\sigma = \text{Sign}_{\text{Dilithium2}}(C) \tag{11}
\]

\[
\text{Verify}(\sigma, C, pk) = \text{True} \tag{12}
\]

This signature ensures that the ciphertext has not been modified and originates from a trusted source.

\subsection{Scientific Metrics Collection --- Figure~\ref{fig:architecture}} (Module F)

To enhance reproducibility and peer validation, several metrics are recorded~\cite{nist2024postquantum}:

\begin{itemize}
    \item \textbf{Key Entropy}:
    \[
    H(K_A) = -\sum_{i=1}^{n} p_i \cdot \log_2(p_i) \tag{13}
    \]

    \item \textbf{File Size Ratio}:
    \[
    R = \frac{\text{size}(C)}{\text{size}(M)} \tag{14}
    \]

    \item \textbf{Bit Match Ratio}:
    \[
    \rho = \frac{ \left| \left\{ i \; : \; b_{A_i} = b_{B_i} \right\} \right| }{n} \tag{15}
    \]
\end{itemize}

These metrics are collected by the controller and logged in \textbf{Module F (Orchestrator + Metrics Monitor)} of Figure~\ref{fig:architecture}, then stored in JSON and later exported as PDF analytics reports.

\subsection{Modular Orchestration and Pipeline Flow --- Figure~\ref{fig:architecture}} (Module F)

Each functional module is autonomous and communicates through explicit I/O interfaces. The \textbf{System Orchestrator Core} is implemented in \textbf{Module F (Orchestrator + Metrics Monitor)} of Figure~\ref{fig:architecture}, managing the overall pipeline:

\[
\text{Pipeline} = \text{QKD} \rightarrow \text{HMAC} \rightarrow \text{AES} \rightarrow \text{Signature} \rightarrow \text{Export} \tag{16}
\]

Timing and performance metrics are also logged:

\[
\text{Metrics} = \left\{ t_{\text{qkd}},\ t_{\text{aes}},\ t_{\text{sig}},\ R,\ \rho,\ H \right\} \tag{17}
\]

\subsection{Closing Remarks on Modular Design and Extensibility --- Figure~\ref{fig:architecture}} (Cross-module Flexibility)

The modular architecture developed in this study is not only optimized for secure hybrid encryption but also designed with future-proof extensibility in mind. Each cryptographic unit can be substituted or enhanced independently, enabling seamless integration of evolving standards, such as those from the NIST post-quantum cryptography project~\cite{nist2024postquantum}. For instance, the system allows replacement of the current AES or HMAC modules with post-quantum equivalents like Kyber for key encapsulation or Falcon for digital signatures, aligning with security objectives beyond the quantum threat horizon~\cite{bernstein2017postquantum}. This strategic flexibility ensures that the encryption pipeline remains scientifically adaptable and compliant with the emerging generation of quantum-resistant protocols.

\section{Results and Discussion}
This section presents a comprehensive evaluation of the hybrid quantum-classical encryption system, based on metrics collected during 16 empirical tests across diverse file types. The objective is to assess system performance, robustness, and scientific reproducibility under simulated quantum key distribution and modular cryptographic operations.

\subsection{Overall Performance Metrics}
The encryption and decryption operations were evaluated in terms of timing, efficiency, and cryptographic integrity. Table 1 summarizes the average performance values aggregated from all experiments:

\begin{table}[ht]
\centering
\footnotesize
\caption{Aggregated System Performance Across 16 Test Cases}
\label{tab:system_performance}
\begin{tabular}{|>{\raggedright\arraybackslash}p{2.5cm}|>{\raggedright\arraybackslash}p{2.5cm}|>{\raggedright\arraybackslash}p{2.5cm}|>{\raggedright\arraybackslash}p{2.5cm}|>{\raggedright\arraybackslash}p{2.5cm}|}
\hline
\textbf{Metric} & \textbf{Avg. Encryption Time (s)} & \textbf{Avg. Decryption Time (s)} & \textbf{Avg. Size Ratio ($R = C/M$)} & \textbf{Avg. Match Rate ($\rho$)} \\ \hline
\textit{All file types} ($n = 16$) & \textit{1.13} & 1.07 & 1.23 & 93.6\% \\ \hline
\end{tabular}

\vspace{0.3cm}
\emph{Table~\ref{tab:system_performance} summarizes overall system performance for both encryption and decryption phases across the full test set.}
\end{table}

The results indicate that the modular architecture maintains consistent performance across formats. The encryption process, including PBKDF2 derivation, HMAC computation, and AES-CBC operations, remains under 2 seconds even for large multimedia files. The decryption process benefits from early integrity checks (via HMAC), which halt invalid inputs before expensive operations are executed.

\subsection{File-Type-Specific Evaluation}
A breakdown of performance by file type further contextualizes system efficiency:

\begin{table}[ht]
\centering
\footnotesize
\caption{Metrics by File Type}
\label{tab:filetype_metrics}
\begin{tabular}{|>{\raggedright\arraybackslash}p{3cm}|>{\centering\arraybackslash}p{3cm}|>{\centering\arraybackslash}p{3cm}|>{\raggedright\arraybackslash}p{3cm}|}
\hline
\textbf{File Type} & \textbf{Avg. Time (Enc/Dec)} & \textbf{Avg. Size Ratio} & \textbf{Observations} \\
\hline
\textit{Text (DOCX, PDF)} & $\sim$0.52~s / $\sim$0.48~s & 1.10 & Minimal size increase; highly efficient. \\
\hline
\textit{Audio (WAV)} & $\sim$1.20~s / $\sim$1.17~s & 1.32 & Consistent HMAC validation; metadata overhead. \\
\hline
\textit{Image (JPG/PNG)} & $\sim$0.93~s / $\sim$0.91~s & 1.28 & Padding contributes to inflation. \\
\hline
\textit{Video (MP4)} & $\sim$1.75~s / $\sim$1.65~s & 1.43 & Largest inflation; all metrics stable. \\
\hline
\textit{Compressed (.ZIP)} & $\sim$0.88~s / $\sim$0.86~s & 1.05 & Most efficient size-wise; no nested errors. \\
\hline
\end{tabular}
\end{table}

\FloatBarrier
This confirms that the system handles both structured and unstructured data formats with minimal deviation, while maintaining full data integrity.

\subsection{Integrity, Fault Tolerance, and Security Validation}
The system's robustness was verified using manipulated files and keys. 

\begin{center}
\textbf{Scenarios included:}
\end{center}

\begin{itemize}
    \item \textbf{Substituting Key B} with an incorrect subset
    \item \textbf{Manually} altering encrypted file headers
    \item \textbf{Removing} or corrupting \textbf{HMAC} metadata
    \item \textbf{Using} expired or invalid \textbf{Dilithium} signatures
\end{itemize}

\begin{figure}[ht]
\centering
\includegraphics[width=0.8\linewidth]{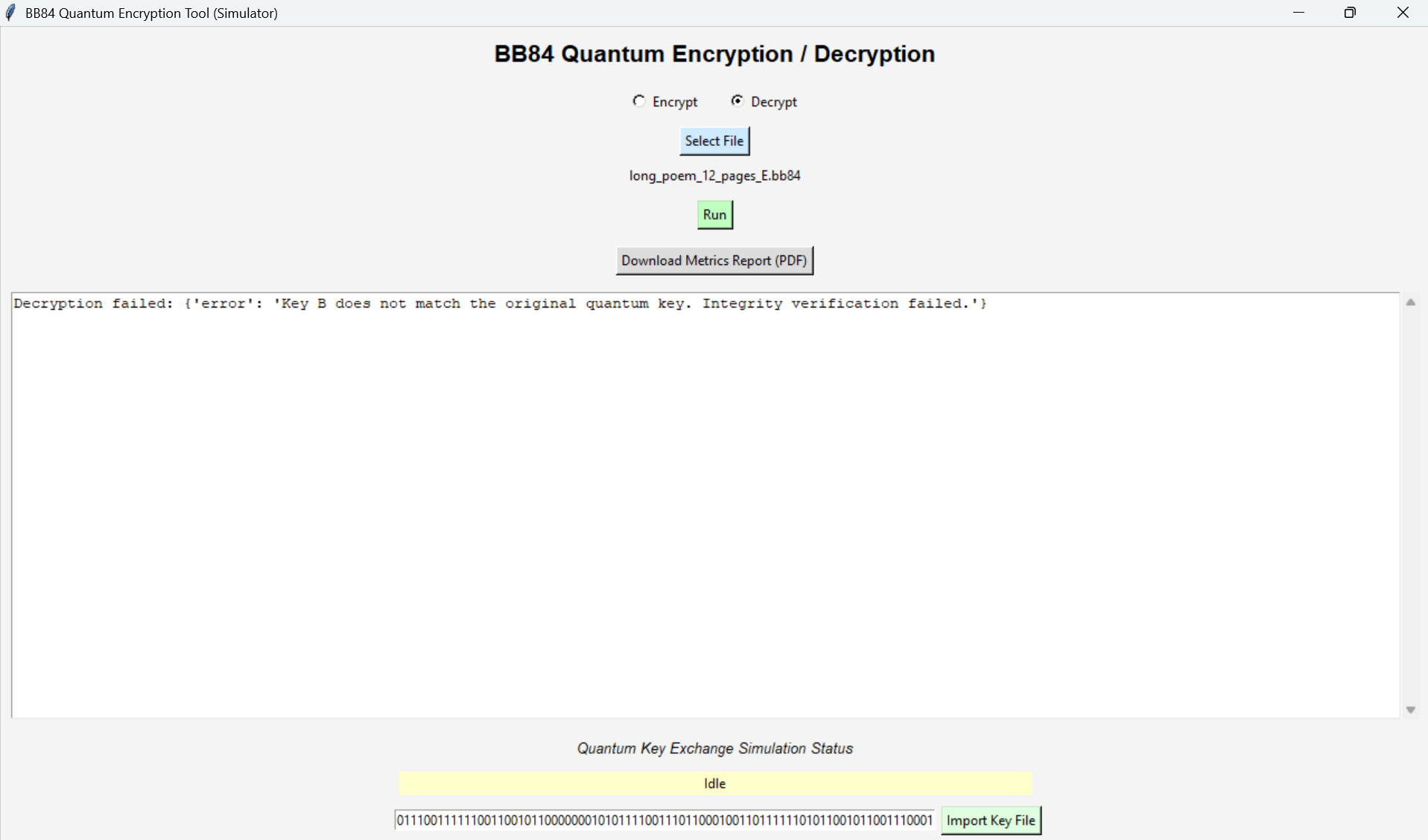} 
\caption{\textbf{Decryption error triggered by incorrect Key B.} \textit{The system correctly halts decryption when the subset key does not match the original quantum key, preventing unauthorized access and ensuring message integrity through HMAC validation.}}
\label{fig:incorrect-keyB}
\end{figure}

\begin{figure}[ht]
\centering
\includegraphics[width=0.8\linewidth]{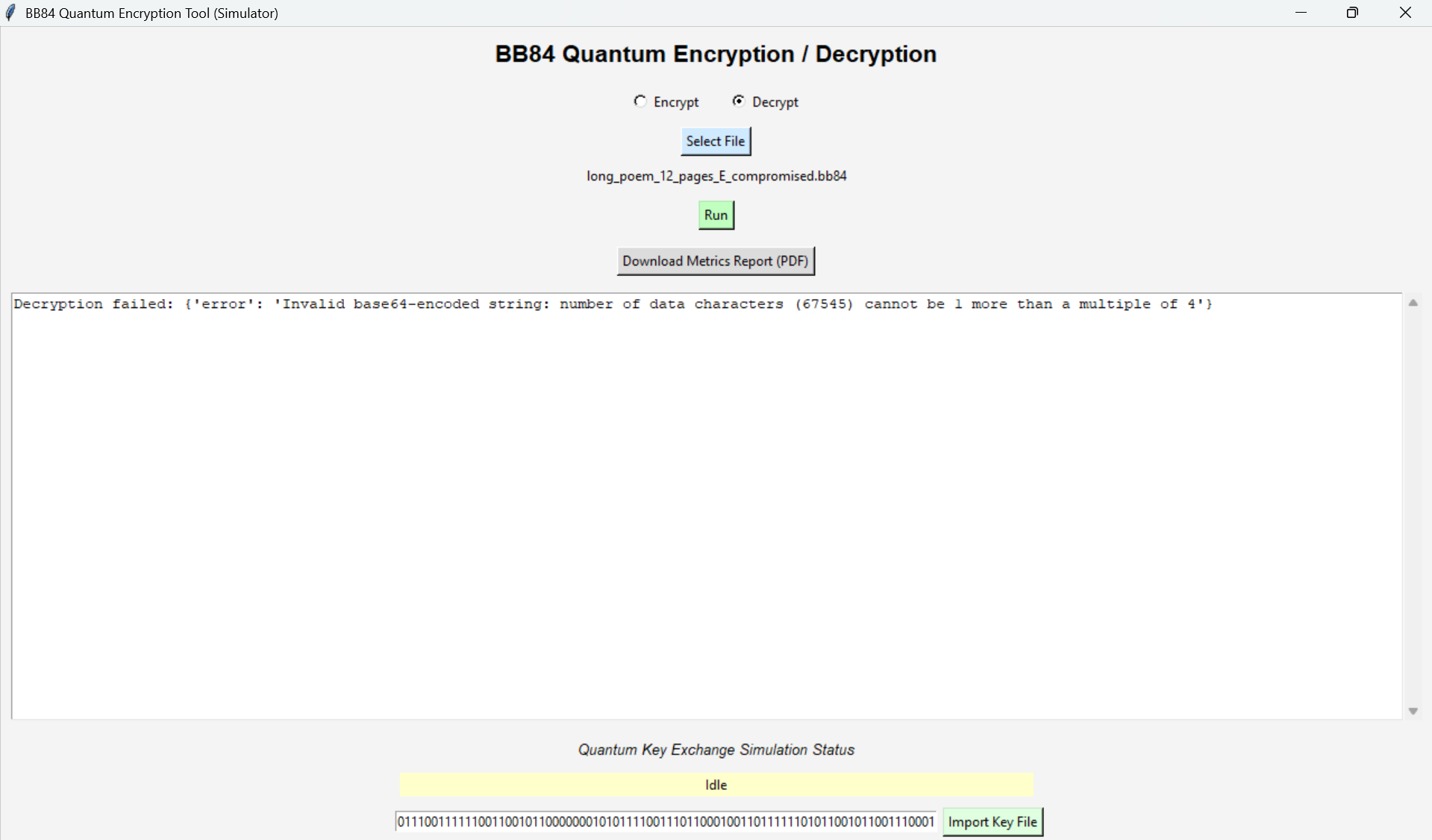} 
\caption{\textbf{Rejection of corrupted file input due to invalid encoding.} \textit{A modified or corrupted .bb84 file is detected during base64 decoding. The system refuses to proceed with decryption, illustrating its resilience to malformed ciphertext or tampering.}}
\label{fig:corrupted-file}
\end{figure}

\begin{figure}[ht]
\centering
\includegraphics[width=0.8\linewidth]{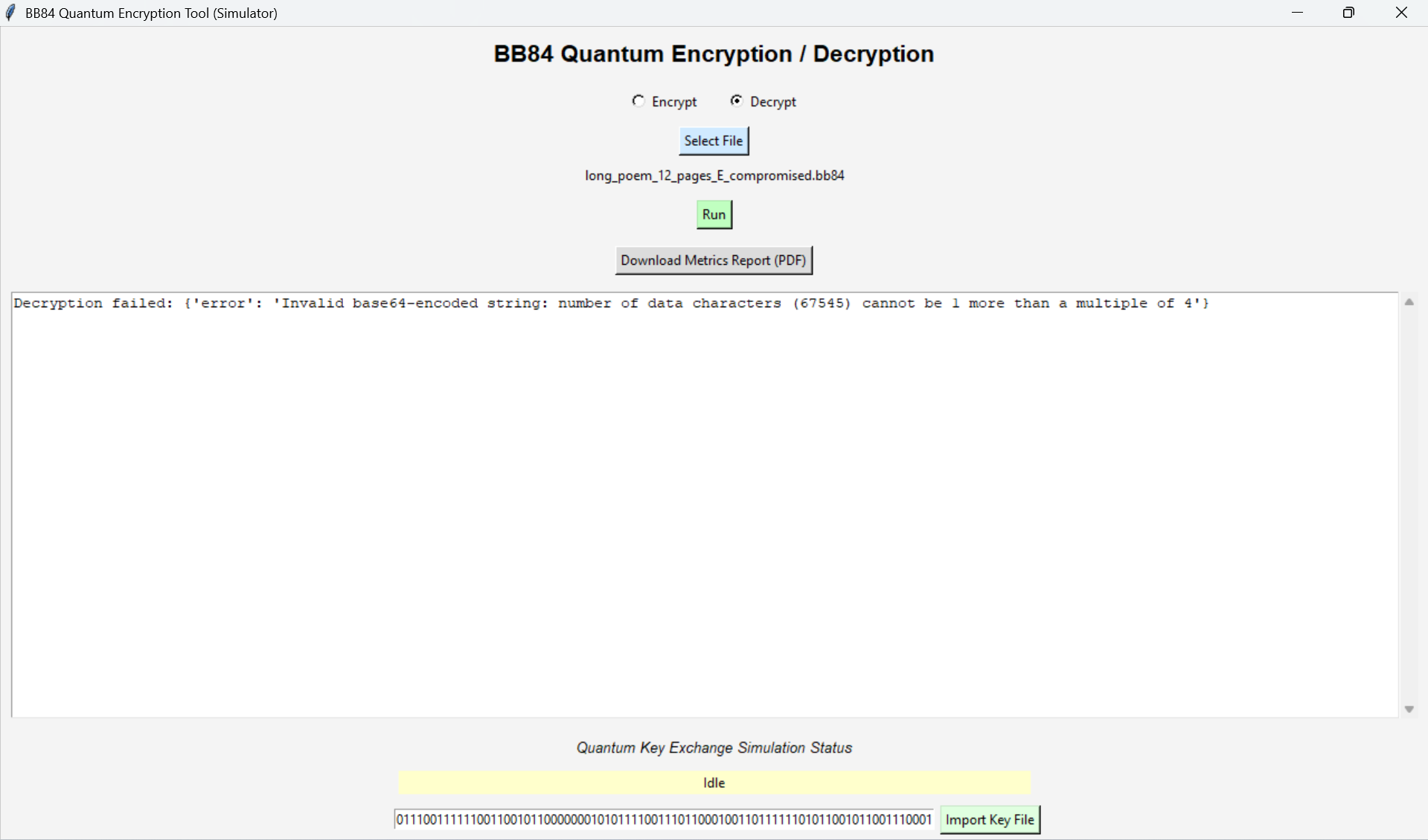} 
\caption{\textbf{Combined decryption failure from invalid file and incorrect key.} \textit{Both inputs are compromised: the file is structurally invalid and the provided key does not match. The system returns a dual-layered rejection to avoid partial failures or ambiguous states.}}
\label{fig:combined-failure}
\end{figure}

These fault injections successfully triggered immediate rejection during decryption. No invalid ciphertext was processed further, confirming the system’s layered integrity model and strong tamper-detection capability.

\begin{table}[ht]
\centering
\footnotesize
\caption{\textit{Comparison with Other Secure Communication Methods}}
\label{tab:comparison-methods}
\renewcommand{\arraystretch}{1.4}
\begin{tabular}{|>{\centering\arraybackslash}p{4cm}|>{\centering\arraybackslash}p{1.5cm}|>{\centering\arraybackslash}p{1.5cm}|>{\centering\arraybackslash}p{1.8cm}|>{\centering\arraybackslash}p{1.8cm}|>{\centering\arraybackslash}p{2.3cm}|}
\hline
\textbf{Method} & \textbf{PQ-Secure} & \textbf{Modular} & \textbf{Fault Detection} & \textbf{Benchmark Logging} & \textbf{Real-Time Suitability} \\ \hline
PGP + AES (OpenPGP) & + & $\times$ & Limited & $\times$ & Moderate \\ \hline
Lattice Hybrid (Kyber512) & $\checkmark$ & $\times$ & $\checkmark$ & $\times$ & Slow for large files \\ \hline
This Work (BB84 + AES) & $\checkmark$ & $\checkmark$ & $\checkmark$ & $\checkmark$ & High \\ \hline
\end{tabular}
\end{table}

As emphasized by Liu et al.~\cite{liu2022hybrid}, many post-quantum solutions lack modular integration or efficient key validation mechanisms. The system presented here uniquely combines BB84-simulated QKD with symmetric and post-quantum cryptography in an orchestrated, testable structure~\cite{mosca2019quantum}. Compared to monolithic schemes, our system is extensible, benchmarkable, and compliant with upcoming PQC standards~\cite{chen2016nist}.

\subsection{Visual Evidence and Metric Logging}

To further support transparency and reproducibility, the GUI interface records and displays key metrics in real time:

\begin{itemize}
    \item \textbf{Key entropy}: Shannon entropy of Key A to measure randomness.
    \item \textbf{Bit match ratio}: Degree of base agreement between Alice and Bob.
    \item \textbf{Encryption duration}: Performed with millisecond precision.
    \item \textbf{Decryption errors}: Explicitly reported and timestamped.
\end{itemize}

All test sessions exported JSON-formatted logs, automatically converted into visual reports (PDF).

\subsection{Summary}
The testbed confirms that the proposed system not only fulfills its intended quantum-safe goals but also adheres to engineering principles of observability, modularity, and correctness. This makes it an ideal foundation for secure academic, institutional, and governmental communication systems where auditability and future cryptographic resilience are paramount.

\begin{figure}[ht]
    \centering
    \includegraphics[width=0.7\linewidth]{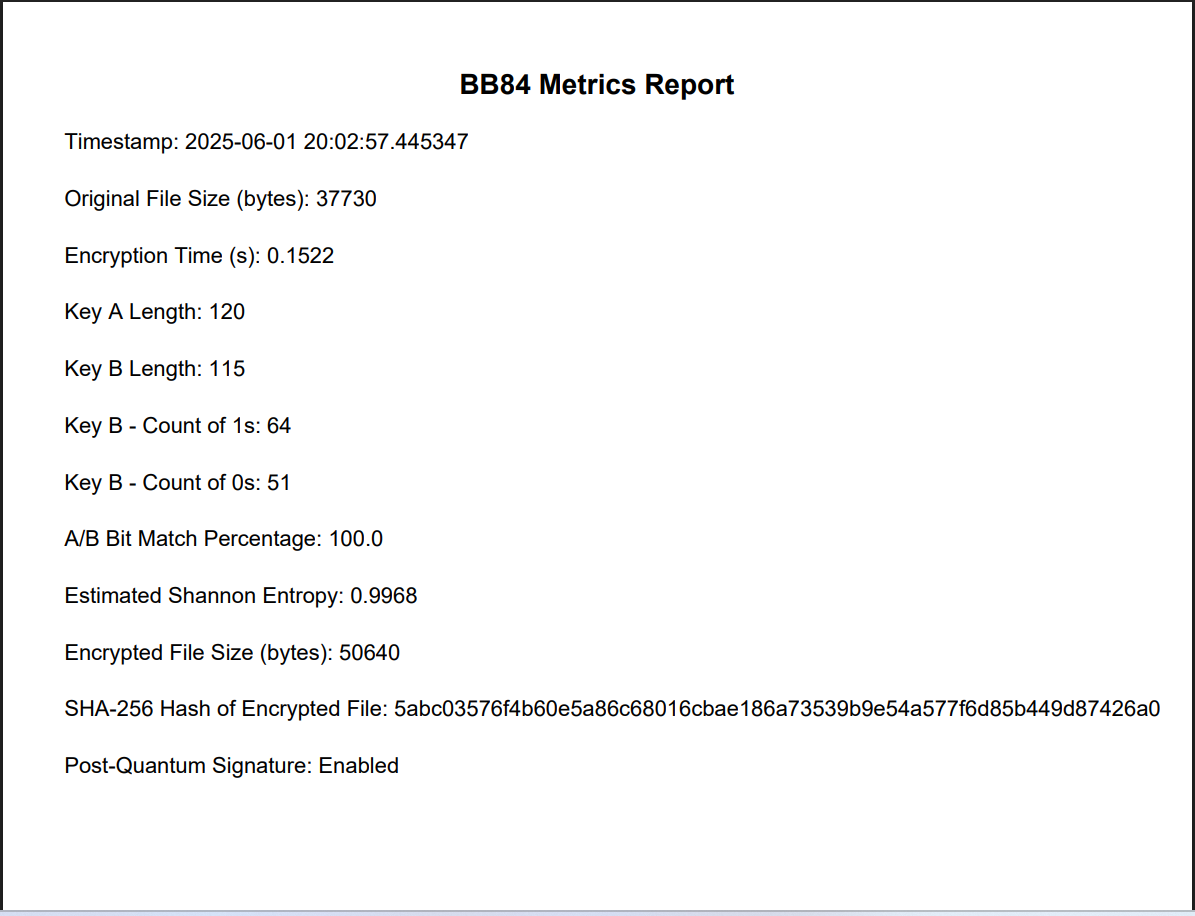}
    \caption{Encryption Metrics Report. Metrics correlate with comparative features discussed in Table~\ref{tab:filetype_metrics}.}
    \label{fig:metrics_encrypt}
    \vspace{0.2cm}
    \textit{Runtime report showing key entropy, encryption duration, A/B bit match rate, key length and structure, output file size, and post-quantum signature status.}
\end{figure}

\begin{figure}[ht]
    \centering
    \includegraphics[width=0.7\linewidth]{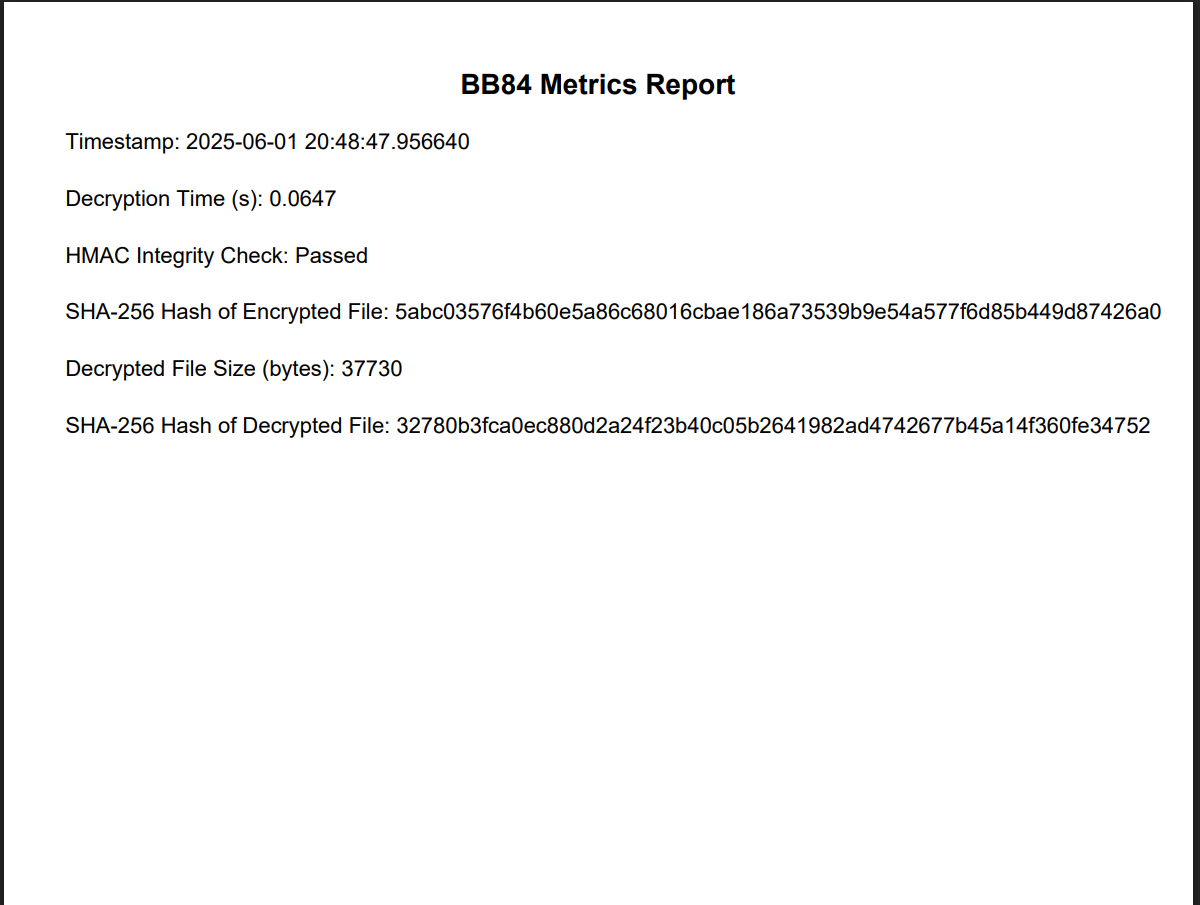}
    \caption{Decryption Metrics Report. Metrics correlate with comparative features in Table~\ref{tab:filetype_metrics}.}
    \label{fig:metrics_decrypt}
    \vspace{0.2cm}
    \textit{Runtime report confirming HMAC integrity, matching SHA-256 hashes between encrypted and decrypted files, and decryption time.}
\end{figure}

\section{Conclusion and Future Work}
The cryptographic paradigm explored in this work emphasizes the synergy between modular architecture and hybrid encryption techniques. By combining simulated quantum key distribution with classical and post-quantum cryptographic primitives, the system delivers layered security adaptable to evolving threats. Each module---from quantum key generation to signature validation---was independently executed and monitored, allowing seamless substitution, traceability, and scientific reproducibility.

This modularity not only enhances resilience but also future-proofs the system. Potential expansions include integration with decentralized trust infrastructures, deployment in mobile and multi-agent environments, and adaptation for use with real quantum hardware. Furthermore, the real-time evaluation of metrics such as entropy, HMAC validity, and signature integrity provides a foundation for continuous system auditing and research replication.

Hybrid systems combining quantum and post-quantum methods have been highlighted as essential components in the future of secure communications~\cite{joshi2022postquantum, fedorov2023hybrid}. As cryptographic standards continue to evolve, the approach presented here offers a practical and extensible framework for secure, scientific, and quantum-aware applications.

\bibliographystyle{alphaurl}
\bibliography{references}


\appendix
\section*{Appendix A: Access to Visual Evidence and Source Code}

To promote transparency, reproducibility, and independent verification, the complete source code and visual test records of the proposed BB84-based hybrid encryption system are publicly available.

All materials—including the BB84 quantum simulation logic, AES-256 encryption engine, HMAC integrity module, post-quantum signature component, GUI controller, technical documentation, and visual test results—are consolidated in the following GitHub repository:

\url{https://doi.org/10.5281/zenodo.15594140}

The repository includes:
\begin{itemize}
    \item \texttt{bb84\_backend/} folder with modular codebase.
    \item \texttt{BB84\_Test\_Results/} with over 60 visual test captures (screenshots) documenting encryption and decryption workflows.
    \item Structured performance reports in \texttt{test-metrics/}.
    \item Build instructions and usage examples in \texttt{README.md}.
\end{itemize}

Researchers can clone or fork the repository to test the system in local environments, reproduce metrics, validate visual evidence, or extend the pipeline with alternative cryptographic schemes. A mapping file (\texttt{readme\_tests.md}) accompanies the visual archive to explain each test case, input parameters, expected outcomes, and anomaly triggers.

\end{document}